\documentstyle[12pt]{article}
\renewcommand \baselinestretch{1.2}

\begin{document}

\hfill UB-ECM-PF 95/11

\hfill April 1995

%\vspace*{1mm}

\begin{center}

{\LARGE \bf Phase structure of renormalizable four-fermion
models in spacetimes of constant curvature}

\vspace{8mm}

\hspace*{-8mm}
{\bf E. Elizalde$^{a,b,}$}\footnote{E-mail:
eli@zeta.ecm.ub.es},
{\bf S. Leseduarte$^{b,}$}\footnote{E-mail:
lese@zeta.ecm.ub.es},
{\bf S.D. Odintsov$^{b,}$}\footnote{On leave of absence from
Tomsk Pedagogical Institute, 634041 Tomsk, Russia.
E-mail: odintsov@ecm.ub.es},
and {\bf Yu.I. Shil'nov$^{c}$}
%\footnote{E-mail: august@ceab.es}
\vspace{4mm}

$^a$Centre for Advanced Studies CEAB, CSIC,
Cam\'{\i} de Santa B\`arbara, 17300 Blanes, Spain \\
$^b$Department ECM and IFAE,
Faculty of Physics, University of  Barcelona, \\
Diagonal 647, 08028 Barcelona, Spain \\
$^c$Department of Theoretical Physics, Kharkov State University, \\
Svobody Sq. 4, Kharkov 310077, Ukraine \\

\vspace{8mm}

{\bf Abstract}
\end{center}

A number of 2d and 3d four-fermion models which are
renormalizable ---in the $1/N$
expansion--- in a maximally symmetric constant curvature space, are
investigated.
To this purpose, a powerful method for the exact
study of spinor heat kernels and propagators on maximally
symmetric spaces is reviewed. The renormalized effective
potential is found for any value of the curvature and
its asymptotic expansion is  given explicitly, both for small and
for strong curvature. The influence of gravity on the dynamical
symmetry breaking pattern of some U(2) flavor-like and discrete
symmetries is described in detail.
%It is seen explicitly that the effect of a
%negative curvature is similar to that of a magnetic field.
The phase
diagram in $S^2$ is constructed and it is shown that, for any
value of the coupling constant, a curvature exists above
which  chiral symmetry is restored. For the case of $H^2$,
 chiral symmetry is always broken. In three dimensions,
in the case of positive curvature, $S^3$, it is seen that
curvature can induce a second-order phase transition.
 For $H^3$ the configuration
given by the auxiliary fields equated to zero is not a solution of
the gap equation.

\vfill
\noindent PACS: 04.62.+v, 04.60.-m, 02.30.+g

\newpage

\section{Introduction}

Four-fermion models \cite{17,1} ---usually considered in the $1/N$
expansion--- are interesting due to the fact that they
provide the opportunity to carry out an explicit, analytical study of
composite bound states and  dynamical chiral symmetry breaking.
At the same time, these theories ---and specially their
renormalizable 2d \cite{1} and 3d \cite{18} variants--- exhibit specific
properties which are similar to the basic behaviors of some
realistic
models of particle physics. Moreover, this class of theories can be used
for the description of the standard model (SM) itself, or of some
particle physics phenomena in the SM (see \cite{19}-\cite{22}).
For example, the dynamical symmetry breking pattern of
Nambu-Jona-Lasinio (NJL) models for the electroweak interaction,
with the top quark as an order parameter, has been discussed in
\cite{19,20}.

Having in mind the applications of four-fermion models to the
early universe and, in particular, the chiral symmetry phase
transitions that take place under the action of the external
gravitational field, there has been recently some activity in the
study of 2d \cite{2}, 3d \cite{13} and 4d \cite{11}-\cite{24}
 four-fermion models in curved spacetime (for a
general introduction to quantum field theory in a curved
spacetime, see \cite{25}). The block-spin renormalization group
(RG) approach and the similarities of the model with the Higgs-Yukawa
one have been considered in \cite{23} and \cite{24}, respectively.

The effective potential of composite fermions in curved spacetime
has been calculated in different dimensions \cite{2}-\cite{12}.
Dynamical chiral symmetry breaking, fermionic mass generation and
curvature-induced phase transitions have been investigated in
full detail. However, in most of these cases only the linear
curvature terms of the effective potential have been taken into
account \cite{13}-\cite{26}. But it turns out in practice that it
is often necessary to consider precisely the strong curvature effects to
dynamical symmetry breaking. In fact we will see that going
beyond the linear-curvature approximation can lead to qualitatively
different results.

In this paper we will investigate some 2d
and 3d four-fermion models which are renormalizable ---in the $1/N$
expansion--- in a maximally symmetric constant-curvature space (either
of positive or of negative curvature). The renormalized effective
potential will be found for any value of the curvature and the
possibility of dynamical symmetry breaking in a curved spacetime
will be carefully explored. Furthermore, the phase structure of the
theory will be described in detail.

The paper is organized as follows. In the next section we
calculate the effective potential of composite fermions in the
Gross-Neveu model, in the spaces $S^2$ and $H^2$.  The phase
diagram in $S^2$ is constructed and it is shown that for any
value of the coupling constant there exists a curvature above
which  chiral symmetry is restored. For the case of $H^2$, we show
that chiral symmetry is always broken. The asymptotic expansions of
the effective potential are given explicitly, both for small and
for strong curvature.
The three-dimensional case is studied in Sect. 3.
We consider two different
four-fermion models: one which exhibits a continuous U(2) symmetry
and another where we concentrate ourselves on two discrete
symmetries which happen never to be simultaneously broken
(see \cite{semenoff}). In Sect. 4 we study explicitly the
dynamical P and Z$_2$ symmetry breaking pattern in $H^3$ and
$S^3$. Finally, Sect. 5 is devoted to conclusions and some technical
points of the procedure are summarized in two appendices.
\bigskip

\section{The Gross-Neveu model in a space of constant curvature}

\subsection{Case of the 2d De Sitter space $S^2$}

In this subsection we will undertake the discussion of the
Gross-Neveu model \cite{1} in De Sitter space. This model,
although rather simple in its conception, displays a quite rich
structure, similar to that of realistic four-dimensional theories
---as renormalizability, asymptotic freedom \cite{1,10} and
dynamical chiral symmetry breaking. Some discussions of chiral
symmetry restoration in the Gross-Neveu model for different
external conditions (such as an electromagnetic field, non-zero
temperature or a change of the fermionic number density) have
appeared in the past \cite{4,7,witten} (the influence of
kink-antikink configurations on the phase transitions is described in
\cite{15,16}).

The study of the Gross-Neveu model in an external gravitational
field has been performed in Ref. \cite{2} using the Schwinger method
\cite{3} (for other analysis of two-dimensional models in curved
space, see \cite{8}). Unfortunately, the generalization of the
Schwinger procedure to curved spacetime is not free from
ambiguities and this is why the result of Ref. \cite{2} includes
some mistake.

In the calculation of the effective potential in the Gross-Neveu
model on De Sitter space we will use a rigorous mathematical
treatment of the fermionic propagator in (constant curvature)
spacetime, that has been developed in Ref. \cite{9}. We shall
start from the action
\begin{equation}
S= \int d^2x \, \sqrt{-g} \left[ \bar{\psi} i \gamma^\mu (x)
\nabla_\mu \psi + \frac{\lambda}{2N} (\bar{\psi}\psi)^2 \right],
\label{2.1}
\end{equation}
where $N$ is the number of fermions, $\lambda$ the coupling
constant, $\gamma^\mu (x) = \gamma^a e^\mu_a (x)$, with $\gamma^a$
the ordinary Dirac matrix in flat space, and $\nabla_\mu$ is the
covariant derivative. By introducing the auxiliary field
$\sigma$, it is convenient to rewrite (\ref{2.1}) as
\begin{equation}
S= \int d^2x \, \sqrt{-g} \left[ \bar{\psi} i \gamma^\mu (x)
\nabla_\mu \psi - \frac{N}{2\lambda} \sigma^2 - \sigma
\bar{\psi}\psi\right],
\label{2.2}
\end{equation}
with $\sigma = - \frac{\lambda}{N} \bar{\psi}\psi$. Furthermore,
in order to apply the results of Ref. \cite{9}, it is convenient
to use Euclidean notations.
%with
%\begin{equation}
%\gamma^0= \gamma^2, \ \ \ \ \gamma^1_E= -i \gamma^1, \ \ \ \  i
%\widehat{\nabla} - - \widehat{\nabla}_E.
%\label{2.3}
%\end{equation}
Then (\ref{2.2}) is written as
\begin{equation}
S= \int d^2x \, \sqrt{g} \left[ \bar{\psi} \gamma^\mu (x)
\nabla_\mu \psi + \frac{N}{2\lambda} \sigma^2 + \sigma
\bar{\psi}\psi\right].
\label{2.4}
\end{equation}

Assuming that we work in De Sitter space and using the standard
$1/N$ expansion, we get the effective potential in terms of
the $\sigma$ field as follows \cite{2,11,12,13}
\begin{equation}
V(\sigma) = \frac{\sigma^2}{2\lambda} + \mbox{Tr }\,  \int_0^\sigma
D(x,x,s)\, ds,
\label{2.5}
\end{equation}
where the propagator $D$ is defined by
\begin{equation}
(\widehat{\nabla} +s) D(x,y,s) = - \delta_2(x,y).
\label{2.6}
\end{equation}
%Within the Eucliden notation, the curvature of a
%constant curvature space can be written in the form
The curvature of both $S^d$ and $H^d$ can be written in the form
\begin{equation}
R=\frac{d(d-1)k}{a^2},
%R= \frac{2k}{a^2},
\label{2.7}
\end{equation}
with $k =1$ for $S^d$ and $k=-1$ for $H^d$; $a$ stands for the radius of
the manifold.
%(anti-De Sitter
%space).

We consider first $k=1$. Following \cite{9} we begin the
calculation by obtaining the `squareing' Green's function:
\begin{equation}
\left(\widehat{\nabla}^2 -s^2\right) G(x,y) = - \delta_2(x,y).
\label{2.8}
\end{equation}
Then $D=\left(\widehat{\nabla} -s\right) G(x,y)$. We resort to the
{\it Ansatz} \cite{9}
\begin{equation}
 G(x,y) = u(x,y) g(p),
\label{2.9}
\end{equation}
where $u(x,y)$ yields a unit matrix in spinor indices
when $y \rightarrow x$, and
$p$ is the distance between $x$ and $y$ along the geodesic that
goes through these two points. Introducing the notations
\begin{equation}
\theta=\frac{p}{a}, \ \ \ \ g(\theta)=\cos  \frac{\theta}{2} \,
h(\theta ), \ \ \ \  Z=\cos^2  \frac{\theta}{2},
\label{2.10}
 \end{equation}
this leads to the following equation for $h$
\begin{equation}
\left[ Z(1-Z) \frac{d^2}{dZ^2} + (2-3Z) \frac{d}{dZ} -1-s^2a^2
\right] h(\theta) =0.
\label{spheq}
\end{equation}
One can show \cite{9} that the solution of (\ref{2.8}) is given by a linear
combination of the hypergeometric functions \cite{14}.
As boundary conditions for (\ref{2.8}) it is convenient to choose
the following. First, one selects the singularity that appears
for $\theta \rightarrow 0$. Second, one demands \cite{9} that the
singular part of this limit has the same form as in flat space
\begin{equation}
g(\theta) \sim - \frac{1}{2\pi} \ln \theta, \ \ \ \theta
\rightarrow 0.
\label{2.12}
\end{equation}
Using properties of the hypergeometric functions and the boundary
conditions (\ref{2.12}), the function $D$ for coinciding
arguments is found to be
\begin{equation}
D(x,x,s) = -s \lim_{\theta \rightarrow 0} g(\theta) =
\frac{s}{4\pi} \left[  \psi (1+isa) + \psi (1-isa)+ 2\gamma + \ln
\frac{\theta^2}{4}\right],
\label{2.13}
\end{equation}
where (\ref{2.12}) has been used explicitly, $\gamma$ is the
Euler constant and $\psi$ the digamma function, and the two arguments
of $D$
are supposed to be separated by a small geodesic distance $p=\theta a$.
Differentiating
(\ref{2.5}) with respect to $\sigma$ and using (\ref{2.13}), one
gets
\begin{eqnarray}
V'(\sigma ) &=& \frac{\sigma}{\lambda} + \mbox{Tr }\,  D(x,x,\sigma) =
\frac{\sigma}{\lambda} \left\{ 1 + \frac{\lambda}{2\pi} \left[
2\gamma + \psi \left( 1 + i|\sigma| \sqrt{\frac{2}{R}} \right)
\right.\right. \nonumber \\
&& \left. \left. +\psi \left( 1 - i|\sigma| \sqrt{\frac{2}{R}}
\right) + \ln \frac{p^2R}{8} \right] \right\},
\label{2.14}
\end{eqnarray}
where $R$ is the curvature.

As renormalization conditions we choose the following \cite{2}
\begin{equation}
\left. V''(\sigma ) \right|_{R=0, \sigma =\mu} =
\frac{1}{\lambda}.%,
\label{2.15}
\end{equation}
%which corresponds to the absence of quantum corrections to the
%mass for $R=0$ at the renormalization point $\mu$.
By selecting the counterterms of the form
\begin{equation}
\delta V = - \frac{1}{2\pi} \sigma^2 \ln \frac{\mu \, p \, e^{\gamma
+1}}{2},
\label{2.15p}
\end{equation}
adding them to (\ref{2.14}) and using the asymptotics of $\psi
(x)$, we find the following value for the derivative of the
renormalized effective potential
\begin{equation}
V' (\sigma) = \frac{\sigma}{\lambda} \left\{ 1
+ \frac{\lambda}{2\pi} \left[ \psi \left( 1 + i|\sigma|
\sqrt{\frac{2}{R}} \right)  +\psi \left( 1 - i|\sigma|
\sqrt{\frac{2}{R}} \right) -2- \ln \frac{2\mu^2}{R} \right] \right\}.
\label{2.16}
\end{equation}
As we can see, letting aside the different notation employed for
$\sigma$, the terms involving $\psi (x)$ differ from the
corresponding terms reported in \cite{2}.

Starting now from expression (\ref{2.16}), different physical
questions can be studied. In particular, the possibility to
construct a corresponding phase diagram appears. To this end, let
us calculate the second derivative
\begin{eqnarray}
V'' (\sigma)& =& \frac{1}{\lambda}
+ \frac{1}{2\pi} \left[  \psi \left( 1 + i|\sigma|
\sqrt{\frac{2}{R}} \right)  +\psi \left( 1 - i|\sigma|
\sqrt{\frac{2}{R}} \right) -2- \ln \frac{2\mu^2}{R} \right]
 \nonumber \\ && + \frac{i|\sigma|}{2\pi} \sqrt{\frac{2}{R}}
\left[\zeta \left(2, 1 + i|\sigma| \sqrt{\frac{2}{R}} \right)  -
\zeta \left(2, 1 - i|\sigma| \sqrt{\frac{2}{R}} \right) \right].
\label{2.17}
\end{eqnarray}
We can now study the behavior of the renormalized $V$ near $\sigma =0$.
We have always
\begin{equation}
V'(0)=0, \ \ \ \ V''(0) = \frac{1}{\lambda} \left[ 1-
\frac{\lambda}{2\pi} \left( 2\gamma + 2+ \ln \frac{2\mu^2}{R}
\right) \right].
\label{2.18}
\end{equation}
With the notation
\begin{equation}
R_0 = 2\mu^2 e^{2(\gamma +1)}, \ \ \ \ \ \ \lambda_0 =2\pi,
\label{2.19}
\end{equation}
we obtain that the point $\sigma =0$ is a minimum for $R>R_0 \exp
(-\lambda_0/\lambda)$ and a maximum for  $R<R_0 \exp (-
\lambda_0/\lambda)$. That is, for any value of $\lambda$, there
exists a value of the curvature above which chiral symmetry
is restored. The connection of the critical curvature with the
%critical
coupling constant is
%This can be shown diagramatically by writting the
%simple equation that connects the critical curvature with the
%critical coupling constant, i.e.
\begin{equation}
R_{cr} = R_0 e^{-\lambda_0/\lambda},
\label{2.20}
\end{equation}
a different expression for $R_{cr}$ will be given below, in
which the independence of  $R_{cr}$ from the renormalization
scale $\mu$ is made apparent.
%and by drawing the phase diagram in the $R-\lambda$ plane (Fig.
%1).
For $R<R_{cr}$ the chiral symmetry is broken and a dynamical
fermion mass is generated.

As next step one can investigate different limits of the
expression
(\ref{2.16}) for the renormalized potential.
For $R \rightarrow 0$, one obtains from
(\ref{2.16})
\begin{eqnarray}
V' (\sigma)& \sim & \frac{\sigma}{\lambda} \left[ 1
+ \frac{\lambda}{2\pi} \left( \ln \frac{\sigma^2}{\mu^2} -2 +
\frac{R}{12 \sigma^2} \right) \right], \label{2.21} \\
V (\sigma)& \sim & \frac{\sigma^2}{2\lambda} \left[ 1
+ \frac{\lambda}{2\pi} \left( \ln \frac{\sigma^2}{\mu^2} -3
\right) \right] + \frac{R}{48 \pi} \ln  \frac{\sigma^2}{R}.
 \label{2.22} \end{eqnarray}
Notice that in (\ref{2.22}) the constant of integration has been
chosen having in mind the finiteness of $V$ as
$R\rightarrow
0$  \cite{2}.
At the same time, at $R \rightarrow \infty$ one gets
%(see also
%\cite{2})
\begin{eqnarray}
V' (\sigma)& \sim & \frac{\sigma}{\lambda} \left[ 1
+ \frac{\lambda}{2\pi} \left(-2\gamma -2 + 4\zeta{(3)}
\frac{\sigma^2}{R} - \ln \frac{2 \mu^2}{R} \right) \right],
\label{2.23} \\
V (\sigma)& \sim & \frac{\sigma^2}{2\lambda} \left[ 1
+ \frac{\lambda}{2\pi} \left(-2\gamma -2 + 2\zeta{(3)}
\frac{\sigma^2}{R} - \ln \frac{2 \mu^2}{R}  \right) \right].
 \label{2.24} \end{eqnarray}
The last expressions show the behavior of the effective potential
at strong curvature. Analysing (\ref{2.21})-(\ref{2.24}) one can
see that for small $R$ chiral symmetry is broken, as it happens
in
flat spacetime. However, in the limit of strong curvature chiral
symmetry is restored. Thus, the study of the asymptotics is a
further check of our general analysis (phase diagrams).

If one defines $M$ by
\[ V''(\sigma ) \left| _{R=0,\,\,\sigma = M} \right. =0 ,\]
then one can express the derivative of the effective potential
in terms of this parameter, as
\[ V' (\sigma ) = \frac{\sigma}{2 \pi} \left[ \psi \left(
1+i\sqrt{\frac{2 \sigma ^2}{R}} \right) +  \psi \left(
1-i\sqrt{\frac{2 \sigma ^2}{R}} \right) + \ln{\frac{R}{2 M^2}}
\right] . \]
We may now rephrase the criterion of symmetry restoration in terms of
$M$ by saying that the symmetry is restored when $R > R_{cr} =
2 M^2 \exp{(2 \gamma)}$. The shape of the effective potential for
different values of the quotient $\frac{R}{M^2}$ is shown in
Fig. 1.
The character of the transition is continuous,
as illustrated by Fig. 2.
%$R\rightarrow \infty$
\medskip

\subsection{Case of the hyperbolic space $H^2$}

In the hyperbolic space $H^2$ (negative curvature) the analysis
can
be carried out in a very similar way. After introducing the
notations
\begin{equation}
\theta=\frac{p}{a}, \ \ \ \ g(\theta)=\cosh \frac{\theta}{2} \,
h(\theta ), \ \ \ \  Z=\cosh^2  \frac{\theta}{2},
\label{2.25}
 \end{equation}
Eq. (\ref{2.8}) acquires the form \cite{9}
\begin{equation}
\left[ Z(1-Z) \frac{d^2}{dZ^2} + (2-3Z) \frac{d}{dZ} -1+s^2a^2
\right] h(\theta) =0.
\label{2.26}
 \end{equation}
As a result, using a similar procedure as in Sect. 2.1, we get
\begin{equation}
D(x,x,s) = \frac{s}{4\pi} \left[ 2\gamma + \psi (1+ sa) + \psi (sa) +
\ln
\frac{\theta^2}{4} \right].
\label{2.27}
 \end{equation}
The non-renormalized expression for $V'(\sigma)$ is
\begin{equation}
V'(\sigma ) = \frac{\sigma}{\lambda} \left\{ 1 +
\frac{\lambda}{2\pi} \left[
2\gamma + \psi \left( 1 +|\sigma|  \sqrt{\frac{2}{|R|}} \right)
 +\psi \left( |\sigma| \sqrt{\frac{2}{|R|}}
\right) + \ln \frac{p^2|R|}{8} \right] \right\}.
\label{2.28}
\end{equation}
Making the same renormalization as in Sect. 2.1, we obtain the renormalized
effective potential
\begin{eqnarray}
V'(\sigma ) &=& \frac{\sigma}{\lambda} \left\{ 1 +
\frac{\lambda}{2\pi} \left[
2\psi \left( 1 + |\sigma| \sqrt{\frac{2}{|R|}} \right) -2 - \ln
\frac{2\mu^2}{|R|} \right] \right\} -\sqrt{\frac{|R|}{8\pi^2}},
\label{2.29} \\
V''(\sigma ) &=& \frac{1}{\lambda} \left\{ 1 +
\frac{\lambda}{2\pi} \left[
2\psi \left( 1 + |\sigma| \sqrt{\frac{2}{|R|}} \right) -2 - \ln
\frac{2\mu^2}{|R|} \right] \right\}  \nonumber \\  && +
\frac{|\sigma|}{\pi}\sqrt{\frac{2}{|R|}} \zeta \left(2, 1+|\sigma|
\sqrt{\frac{2}{|R|}} \right).
\label{2.30}
\end{eqnarray}
A careful study of $V'(0)$ shows that due to the presence of the
last
term in (\ref{2.29}), $\sigma=0$ is never
%an extremum,
stationary
for any value
of
$\lambda$ and finite $R$.
Owing to the fact that $V'(0)
<0$,
chiral symmetry is always broken in $H^2$.
%chiral symmetry is broken at $\sigma =0$, and (\ref{2.20}) points
%out to the values of $R$ and $\lambda$ at which the slope of
%$V (0)$ at $\sigma =0$ is changing.

 In the small curvature
limit ($|R|\rightarrow 0$)
\begin{equation}
V(\sigma ) = \frac{\sigma^2}{2\lambda} \left[ 1+
\frac{\lambda}{2\pi} \left(\ln \frac{\sigma^2}{\mu^2}-3 \right) \right] -
\frac{|R|}{48\pi} \ln \frac{\sigma^2}{|R|} ,
\label{2.31}
\end{equation}
what coincides with (\ref{2.22}), taking into account the change
of
sign for the curvature.
In the strong curvature limit ($|R|\rightarrow \infty$)
\begin{equation}
V(\sigma ) = \frac{\sigma^2}{2\lambda} \left[ 1-
\frac{\lambda}{2\pi} \left(2\gamma +2 + \ln \frac{2\mu^2}{|R|} +
\frac{ \sqrt{2|R|}}{|\sigma|} \right) \right].
\label{2.32}
\end{equation}
The analysis of Eqs. (\ref{2.31}) and (\ref{2.32}) shows that the
general conclusion about the chiral symmetry breaking at any
finite
$R$ in $H^2$ is correct. In a similar way one can study the influence of
curvature in the massive Gross-Neveu model (for a recent
discussion of such model at non-zero temperature, see \cite{casal}).

%This finishes our discussion of the Gross-Neveu model at constant
%curvature. To be noticed is the fact that the conclusion about
%the
%restoration of the broken chiral symmetry at $R=R_{cr}$ in $S^2$
%is
%obtained in general terms without having to resort to asymptotic
%expansions of the effective potential. However, in order to study
%the character of the phase transition, one has to investigate the
%behavior of the potential numerically.
\bigskip

\section{Dynamical U(2) flavor symmetry breaking in $H^3$ and
$S^3$}

This section is devoted to the description of a
three-dimensional four-fermion model
which has a continuous flavor-like symmetry and  how its
breaking is affected by a gravitational background.

We consider the following model on a Riemannian manifold
\begin{equation}
{\cal L}_E = \bar{\psi} \not{\!\! D} \psi - \frac{\lambda_B}{2N}
\left[
(\bar{\psi} \psi)^2 + (\bar{\psi} i \tau^1 \psi)^2 + (\bar{\psi}
i
\gamma^5 \psi)^2 \right],
\end{equation}
with
 \begin{equation}
\gamma^\mu = \left( \begin{array}{cc} \sigma^\mu & 0 \\ 0 & -\sigma^\mu
\end{array}
\right), \ \ \mu = 1,2,3, \ \
\gamma^5 =i \left( \begin{array}{cc} 0 & 1\!\!1_2 \\ - 1\!\!1_2 & 0 \end{array}
\right) =-
\tau^1\gamma^1\gamma^2\gamma^3, \
\ \ \
\tau^1 = \left( \begin{array}{cc}  0& 1\!\!1_2  \\ 1\!\!1_2 & 0 \end{array}
\right), \end{equation}
so we take a reducible, four-dimensional Dirac algebra.

The transformation of the bilinear terms $\bar{\psi} \psi$,
$\bar{\psi} i \tau^1 \psi$ and $\bar{\psi} i\gamma^5 \psi$ under
\begin{equation}
\delta \psi = -i T_\alpha\theta^\alpha \psi, \ \ \ \ \
\delta \bar{\psi} = i \bar{\psi} \gamma^3 T_\alpha \gamma^3
\theta^\alpha,
\end{equation}
with $T_0=1$, $T_1=\gamma^5$, $T_2=\tau^1$ and $T_3=i\tau^1
\gamma^5$, is given by
\begin{equation}
\left( \begin{array}{c} \delta (\bar{\psi} \psi) \\ \delta (\bar{\psi} i
\tau^1 \psi) \\ \delta (\bar{\psi} i\gamma^5 \psi) \end{array} \right) =
2 \left( \begin{array}{ccc} 0 & -\theta^1 & - \theta^2 \\ \theta^1 & 0 & -
\theta^3 \\ \theta^2 & \theta^3 & 0 \end{array} \right)
\left( \begin{array}{c} \bar{\psi} \psi \\ \bar{\psi} i \tau^1 \psi \\
\bar{\psi} i\gamma^5 \psi \end{array} \right).
\end{equation}
It is convenient  to express this theory in terms of auxiliary
fields, namely
\begin{equation}
{\cal L}_E = \bar{\psi} \left( \not{\!\! D} + \sigma + i \tau^1\rho +
i\gamma^5 \pi \right) \psi + \frac{N}{2\lambda_B} (\sigma^2 +
\rho^2+ \pi^2 ),
\end{equation}
what yields
\begin{equation}
\Gamma_{N \rightarrow \infty} [\sigma,\rho, \pi] = \int dx \,
\sqrt{g} \, \frac{N}{2\lambda_B} (\sigma^2 + \rho^2+ \pi^2 ) -
N \ln \det \left( \not{\!\! D} + \sigma + i \tau^1\rho + i\gamma^5
\pi \right).
\end{equation}
We consider the constant configurations of $\sigma, \rho, \pi$.
The rotational symetries impose that the effective potential only
depends on $\sigma^2+\rho^2+\pi^2$, what allows us to set
$\rho=\pi=0$. The regularized expression of the effective
potential  turns out to be:
\begin{equation}
\int dx \, \sqrt{g} \, V [\sigma]
= \int dx \, \sqrt{g} \, \frac{N}{2\lambda_B} \sigma^2 +
\frac{N}{2}  \mbox{Tr }\,  \int_{1/\Lambda^2}^\infty \frac{dt}{t} \, e^{-
t\sigma^2} e^{\not\nabla^2t}.
\end{equation}
The reader may verify that the use of point-splitting regularization leads
to the same results that we get below by regulating the lower limit in
the proper time integral.
On the right hand side we observe the appearance of the
coincidence limit of the heat kernel. Resorting to the results of
appendix B, the outcome is (for $H^3$)
\begin{equation}
 V [\sigma]
=  \frac{N}{2\lambda_B} \sigma^2 + \frac{N}{2\pi^{3/2}}
\int_{1/\Lambda^2}^\infty \frac{dt}{t^{5/2}} \, \left( 1+
\frac{1}{2t} \right) e^{-t\sigma^2} e^{\not\nabla^2t}.
\end{equation}
The dependence on the radius of the manifold is not
shown, but it can very easily be recovered from dimensional
analysis. We obtain
\begin{equation}
\frac{1}{N} V [\sigma]
=  \frac{ \sigma^2}{2\lambda_B} + \frac{|\sigma|}{(4\pi)^{3/2}a^2}
\left[  \Gamma \left( - \frac{1}{2}, \frac{\sigma^2}{\Lambda^2}
\right)+ 2a^2\sigma^2 \Gamma \left( - \frac{3}{2},
\frac{\sigma^2}{\Lambda^2} \right) \right].
\end{equation}
Imposing the renormalization condition
\begin{equation}
\frac{1}{N} \frac{d^2}{d\sigma^2} V [\sigma =0, R=-
6/a^2=0]=  \frac{1}{\lambda},
\label{rc1}
\end{equation}
one gets
\begin{equation}
\frac{1}{\lambda_B} = \frac{1}{\lambda} +
\frac{\Lambda} {\pi^{3/2}}, \label{aa1}
\end{equation}
and in the limit $\Lambda \rightarrow \infty$:
\begin{equation}
\frac{1}{N} V [\sigma]
=  \frac{ \sigma^2}{2\lambda} + \frac{|\sigma|}{\pi}
\left( \frac{\sigma^2}{3}-  \frac{1}{4a^2} \right).
\label{h3pot}
\end{equation}
 From this one can easily see that the symmetry is always broken in $H^3$.
In fact we can also notice the curious feature that the origin is not a
solution of the gap equation. This appears to liken the situation of
four-fermion models in three dimensions under the influence of a
magnetic field (see \cite{naftu}).

As for $S^3$, we have (see appendix B).
\begin{equation}
V(\sigma)=\frac{\sigma^2}{2 \lambda_B} - \frac{1}{2\,\,\,Vol} {\bf Tr}
\int_0^{\sigma^2}
d\,\,s \frac{1}{s-\not\!\partial^2}\,\,. \label{primit}
\end{equation}
 From this formula one derives the regularized effective potential
using the result from \cite{9} that on $S^N$ the solution of
\[ (\not\!\nabla^2-m^2)G(y)=-\delta_N(y) \]
is given by $G(y)=U(y) g_N(\sigma)$, where U(y) is a parallel transport
matrix that is the identity at the coincidence limit, and
\[g_N (\theta)=\frac{\Gamma\left( \frac{N}{2}+im \right)
               \Gamma\left( \frac{N}{2}-im \right)}{ (4 \pi)^{N/2}
               \Gamma\left( \frac{N}{2}+1 \right)}
               \cos \left( \frac{\theta}{2} \right)
               F\left( \frac{N}{2}+im,\,\,\frac{N}{2}-im,\,\,
               \frac{N}{2}+1,\,\,\cos^2 \left( \frac{\theta}{2} \right)
                \right)\,\,,\]
where $\theta = \frac{\rho}{a}$, ($\rho$ is the geodesic distance, and a
is the radius of the manifold). With this in mind and using known
properties of the hypergeometric functions one may arrive at the
following expression (in which we only keep terms divergent in $\rho$
and terms independent from it)
\begin{eqnarray}
V'(\sigma)&=&\frac{\sigma}{\lambda_B} - \frac{\sigma}{4 \pi}
\frac{\mbox{Tr }\,\,1\!\!1_4}{\rho} +
 \frac{\mbox{Tr }\,\,1\!\!1_4}{4 \pi}
\left( \frac{1}{4 a^2}+\sigma^2 \right) \tanh \left( \pi \sigma a
\right)\,\,.
\label{pots3}
\end{eqnarray}
Using the renormalization given by
\[ \frac{1}{N} \frac{d^2}{d\sigma^2} V [\sigma =0, R=
6/a^2=0]=  \frac{1}{\lambda}, \]
we get
\begin{equation}
V'(\sigma)=\frac{\sigma}{\lambda}+
 \frac{1}{\pi}
\left( \frac{1}{4 a^2}+\sigma^2 \right) \tanh \left( \pi \sigma a
\right)\,\,.    \label{final}
\end{equation}

This appears to be in agreement with the correspondent effective potential
in the
Gross-Neveu model (in the simplest version
without continuous $U(2)$ symmetry and discrete
symmetries) on De Sitter space using, different from our approach,
dimensional regularization (see \cite{muka}).
We can now compare this result with the one that was found in Ref. \cite{13},
which was a study of three-dimensional theories in the small curvature
limit. One can easily check that expression (\ref{pots3}) is
indeed compatible with those in \cite{13}. What is most surprising
in the weak curvature limit is that, in view of the results,
 one might be tempted to conclude that
the origin is not a solution to the gap equation (as it happens in $H^3$),
and, furthermore, that there may be no solution at all.
However, looking at the exact result (\ref{pots3}) or (\ref{final}),
 this is
seen to be an artifact of the approximation. The effective potential
given in expression (\ref{final}) may give rise to a second order phase
transition, as we illustrate in Fig 3.

\section{Dynamical P and Z$_2$ symmetry breaking in $H^3$ and
$S^3$}

In this section we analyse a model which displays two discrete
symmetries. First of all, we present the model and later we describe its
symmetries in some detail. To finish, we will describe the influence of gravity
on the breaking of the symmetries.
Using the representation for the $\gamma^\mu$ (which has no $\gamma_5$)
\begin{equation}
\gamma^\mu = \left( \begin{array}{cc} \sigma^\mu & 0 \\ 0 & \sigma^\mu
\end{array} \right),
\end{equation}
it is also possible in this case to define a parity operation
which admits the presence of a mass term, i.e.
\begin{equation}
{\cal L}_E = \bar{\psi} \not\!\partial \psi + m \bar{\psi} \tau^3
\psi,
\end{equation}
with
\begin{equation}
P[ \psi (x_1,x_2,x_3)]= \gamma^2 \tau^1 \psi (x_1,-x_2,x_3),
\qquad P[ \bar{\psi} (x_1,x_2,x_3)]=- \bar{\psi} (x_1,-x_2,x_3)
\gamma^1 \tau^1 ].
\end{equation}
However, $\bar{\psi}\psi$ transforms as a pseudoscalar. There is
another discrete symmetry of the kinetic term, given by
\begin{equation}
Z_2[\psi] = \tau^1\psi, \qquad Z_2[\bar{\psi}] =\bar{\psi}
\tau^1.
\end{equation}
Under this operation, the roles of the mass terms are reversed, e.g.
\begin{equation}
Z_2[\bar{\psi}\tau^3 \psi] = -\bar{\psi}\tau^3\psi, \qquad
Z_2[\bar{\psi}\psi] = \bar{\psi}\psi.
\end{equation}
With this in mind, it is immediate that
\begin{equation}
{\cal L}_E = \bar{\psi} \not{\!\! D} \psi - \frac{\lambda_B}{2N}
(\bar{\psi}\psi)^2 -  \frac{\kappa_B}{2N}(\bar{\psi} \tau^3
\psi)^2
\end{equation}
is invariant under both P and Z$_2$. But, as $\bar{\psi} \tau^3
\psi$ is {\it not} invariant under Z$_2$, it can be taken as an
order parameter for the Z$_2$ symmetry breaking. Likewise,
$<\bar{\psi} \psi>$ is an order parameter for the P symmetry
breaking.

In terms of auxiliary fields
\begin{equation}
{\cal L}_E = \bar{\psi} (\not{\!\! D} + \phi + \chi \tau^3) \psi +
\frac{N}{2\lambda_B} \phi^2 +  \frac{N}{2\kappa_B} \chi^2.
\end{equation}
Proceeding along the same lines we have trodden before, we will
impose equivalent renormalization conditions
\begin{equation}
\frac{d^2}{d \phi^2} V [\phi=0,R=0] =
\frac{1}{\lambda}, \qquad \frac{d^2}{d \chi^2} V
[\chi=0,R=0] = \frac{1}{\kappa}.
\end{equation}
In the case of hyperbolic space $H^3$ we get that the effective
potential in the large-$N$ limit is given by
\begin{eqnarray}
\frac{1}{N} V [\phi, \chi]
&=&  \frac{ \sigma_+^2+ \sigma_-^2}{8} \left(
\frac{1}{\lambda} +\frac{1}{\kappa}\right) +  \frac{
\sigma_+ \sigma_-}{4} \left( \frac{1}{\lambda} -
\frac{1}{\kappa}\right)  \nonumber \\
&& + \frac{1}{2\pi} \left[|\sigma_+| \left(
\frac{\sigma_+^2}{3}-  \frac{1}{4a^2} \right) + |\sigma_-
| \left( \frac{\sigma_-^2}{3}-  \frac{1}{4a^2} \right)
\right],
\end{eqnarray}
with $\sigma_\pm \equiv \phi \pm \chi$. We obtain the two cases
\begin{equation}
\frac{1}{\lambda} -\frac{1}{\kappa} >0, \ \
\bar{\sigma}_+ =-\bar{\sigma}_-, \ \ (\bar{\phi} =0, \bar{\chi}
\neq 0),
\end{equation}
for the Z$_2$ symmetry breaking, and
\begin{equation}
\frac{1}{\lambda} -\frac{1}{\kappa} <0, \ \
\bar{\sigma}_+ =\bar{\sigma}_-, \ \ (\bar{\phi} \neq 0,
\bar{\chi}= 0),
\end{equation}
for the P symmetry breaking,  respectively. This is illustrated in Figs.
4 and 5, where we only consider positive values of $\phi$, as the symmetry
of the model allows us to reproduce the result for the region of
negative $\phi$ immediately. It is also worth noting that
 either P or Z$_2$ is broken,
but that it is impossible to have {\it both} symmetries broken.
 Fig. 4 exemplifies the
first situation (2.68): one sees that the minimum lies at $\phi =0$,
 $\chi
\neq 0$ and that Z$_2$ is broken.
 In Fig. 5 we are in the second situation:
the minimum lies at $\phi \neq  0$, $\chi = 0$ and P is broken.

As for the $S^3$ case, the reader may verify ---along the same lines as
above--- that one encounters again two different situations: either both
$P$ and $Z_2$ are unbroken or just one of them is broken. This is
obtained without difficulty by repeating the same analysis, and we feel
that further details are not necessary.

\bigskip

\section{Conclusions}

In this paper we have reviewed a powerful method for the exact
study of spinor heat kernels and propagators on maximally
symmetric spaces. We have used it with success in a number of
different four-fermion models.
 The renormalized effective
potential has been found,  in each case, for any value of the curvature,
and its asymptotic expansion has been given explicitly, both for small
and for strong curvature.

We have described in great detail the influence of gravity on the
dynamical
symmetry breaking pattern of some U(2) flavor-like and discrete
symmetries.
In particular, we have seen explicitly that the effect of a
negative curvature is similar to that of the presence of a magnetic field.

In the case of positive curvature, $S^3$, we have checked  that
the scenario given in the framework of a small curvature expansion is
dramatically changed when we treat gravity exactly. In particular
we find that the character of the phase transitions induced by
gravity is continuous and that the origin is always a solution of
the gap equation.
A point to be duely remarked is the fact that the techniques applied
here to two- and three-dimensional models work equally well in four- and
higher-dimensional ones, as well. Of course, these models are not
renormalizable in the standard way (see however, refs. \cite{gural} where
mean field renormalization of four-fermion models has been discussed),
but one can still apply the above method in order to
obtain the cut-off dependent effective potential, what can certainly
 be useful
for cosmological applications.

Another quite interesting topic is to study a formulation
overlapping technique between the Schwinger-Dyson equations in a
constant-curvature space and the effective potential approach.
We hope to address some of these questions in the near future.

\vspace{5mm}

%\newpage

\noindent{\large \bf Acknowledgments}

SL gratefully acknowledges an FPI grant from Generalitat de Catalunya.
YIS is indebted with Dr. A. Letwin for kind support.
This work has been partially financed by DGICYT (Spain), projects
PB93-0035 and SAB93-0024, and by RFFR (Russia), project 94-02-03234.
\bigskip \bigskip

\appendix

\section{ Appendix: Generalities about spinors in three dimensions}

Some general facts about spinors in three dimensions that may be
interesting to recall for the benefit of the reader are the
following. The Dirac algebra $\{ \gamma_\mu, \gamma_\nu \} = 2
g_{\mu
\nu}$ has two irreducible representations, namely
\begin{enumerate}
\item $\gamma^0 = \sigma^3, \ \gamma^1 = i \sigma^1, \ \gamma^2 =
i\sigma^2$ \\ and
\item $\gamma_\mu' = - \gamma_\mu$.
\end{enumerate}
On the other hand, there is no `$\gamma^5$' matrix in order 2
which
anticommutes with all of the $\gamma^\mu$s. A mass term of the
form $m
\bar{\psi} \psi$ in the Lagrangian explicitly violates parity,
defined
by
\begin{equation}
P: \psi (x^0,x^1,x^2) \longrightarrow \sigma^1 \psi
(x^0,-x^1,x^2).
\end{equation}
However, if one uses a reducible representation of the Dirac
matrices,
such as
\begin{equation}
\gamma^0 = \left( \begin{array}{cc} \sigma^3 & 0 \\ 0 & -\sigma^3 \end{array}
\right), \
\ \ \
\gamma^1 = \left( \begin{array}{cc} i\sigma^1 & 0 \\ 0 & -i\sigma^1 \end{array}
\right),
\ \ \ \
\gamma^2 = \left( \begin{array}{cc} i\sigma^2 & 0 \\ 0 & -i\sigma^2 \end{array}
\right),
\end{equation}
then the mass term $m\bar{\psi} \psi$ does preserve parity, as
defined
by
\begin{equation}
P: \psi (x^0,x^1,x^2) \longrightarrow \tau^1 \gamma^1 \psi
(x^0,-x^1,x^2), \end{equation}
being
\begin{equation}
\tau^1 = \left( \begin{array}{cc} 0 & 1\!\!1_2 \\ 1\!\!1_2 & 0 \end{array}
\right), \ \ \ \
\tau^2 = \left( \begin{array}{cc} 0 & -i 1\!\!1_2 \\ i 1\!\!1_2 & 0 \end{array}
\right), \
\ \ \
\tau^3 = \left( \begin{array}{cc} 1\!\!1_2 & 0 \\ 0 & -1\!\!1_2 \end{array}
\right). \end{equation}

The kinetic term is invariant under the U(2) transformations with
generators $T_0 =1\!\!1_4, \ T_1 = \gamma^5, \ T_2 = \tau^1$, and $T_3
= i
\tau^1\gamma^5$, where
\begin{equation}
\gamma^5 = \tau^1 \gamma^0 \gamma^1 \gamma^2 = i
 \left( \begin{array}{cc} 0 & 1\!\!1_2 \\ -1\!\!1_2 & 0 \end{array} \right).
\end{equation}
The mass term breaks this symmetry down to U(1) $\times $ U(1)
with the
generators $T_0$, $T_3$.
\medskip

\section{Appendix: Spinor heat-kernel in maximally symmetric spaces (the
intertwining method)}

In this appendix we review a powerful method to derive closed expressions
for the spinor heat-kernel and propagators in maximally symmetric spaces
(for more details see \cite{9}).

The spinor heat kernel satisfies $ \left( -
\frac{\partial}{\partial t}
+ \not\!\nabla^2 \right) K(y,y_0;t) =0$ and the initial condition
$
\lim_{t \rightarrow 0} $ $ K(y,y_0;t) = 1 \ \delta_N (y,y_0)$.
Substituting
the {\it Ansatz} $K(y,t) = U(y) f(\sigma, t)$ in the heat-kernel
equation, one gets
\begin{equation}
-U \frac{\partial f}{\partial t} + U \nabla^a \nabla_a f + 2 n^a
(\nabla_a U) \frac{\partial f}{\partial \sigma}+ (\nabla^a
\nabla_a
U) f -\frac{R}{4} Uf =0,
\end{equation}
where $n_a = \nabla_a \sigma$. Here the term linear in $\partial
f/ \partial \sigma$ and $\nabla_a U$ cancels out, provided that
$U$
satisfies  the parallel transport equation
\begin{equation}
n^a\nabla_a U=0, \qquad U(y_0)=1.
\end{equation}
The Laplacian acting on $f$ can be replaced by its radial part
given by
\begin{equation}
 \nabla^a \nabla_a f = \Box_N f = (\partial_\sigma^2 + (N-1) B
\partial_\sigma )f, \qquad B = \left\{ \begin{array}{ll} \frac{1}{a} \cot
\left( \frac{\sigma}{a}\right), & S^N, \\  \frac{1}{a} \coth \left(
\frac{\sigma}{a}\right), & H^N, \end{array} \right.
\end{equation}
and the Laplacian acting on $U$ is
\begin{equation}
 \nabla^a \nabla_a U =-\frac{1}{4} A^2(N-1)U, \qquad A = \left\{
\begin{array}{ll} - \frac{1}{a} \tan \left( \frac{\sigma}{2a}\right), & S^N, \\
\frac{1}{a} \tanh \left( \frac{\sigma}{2a}\right), & H^N, \end{array} \right.
\end{equation}
The equation for the scalar $f$ becomes
\begin{equation}
(-\partial_t + L_N )f =0, \qquad L_N=\Box_N - \frac{R}{4} -
\frac{1}{4} (N-1)A^2.
\end{equation}
The idea is now to relate the solutions of this equation for
different $N$s. To this end one looks for an operator $D$ such
that
$L_ND=DL_{N-2}$. A simple {\it Ansatz} leads to
\begin{equation}
D = \left\{ \begin{array}{lll}  \frac{1}{2\pi} \cos \frac{\theta}{2}
\frac{\partial}{\partial(\cos \theta)} \, \left( \cos
\frac{\theta}{2} \right)^{-1}, & \theta = \frac{\sigma}{a}, &
S^N,
\\ - \frac{1}{2\pi} \cosh \frac{x}{2} \frac{\partial}{\partial(\cosh
x)}
\, \left( \cosh\frac{x}{2} \right)^{-1}, & x =i \frac{\sigma}{a},
& H^N, \end{array} \right.
\end{equation}
The odd-dimensional case is elementary, taking into account that
$L_ND^{(N-1)/2}=D^{(N-1)/2}L_1$ and the known solution of $(-
\partial_t + L_1)K_1=0$. This yields
\begin{eqnarray}
K_N(y,t) &=& U(y)  \cosh \frac{x}{2} \left(- \frac{1}{2\pi}
\frac{\partial}{\partial(\cosh x)}\right)^{(N-1)/2} \left(
\cosh\frac{x}{2} \right)^{-1} \frac{e^{-x^2/(4t)}}{\sqrt{4\pi
t}},
\ \ H^N, \nonumber \\
K_N(y,t) &=& U(y) \sum_{n=-\infty}^\infty (-1)^n f_{dp} (\theta +
2\pi n,t), \\ && \hspace{-1cm}  f_{dp} (\theta + 2\pi n,t) =
 \cosh \frac{\theta}{2} \left( \frac{1}{2\pi}
\frac{\partial}{\partial(\cosh \theta)}\right)^{(N-1)/2} \left(
\cosh\frac{\theta}{2} \right)^{-1} \frac{e^{-
\theta^2/(4t)}}{\sqrt{4\pi t}}, \ \ S^N. \nonumber
\end{eqnarray}
In the particular case $N=3$ these expressions read ---in the
coincidence limit--- as follows
\[\lim_{y \rightarrow y_0} K(y,t) = 1\!\!1_d \frac{1}{(4 \pi t)^{
3/2}}  \left[1+\frac{t}{2 a^2} \right], \ \ H^3, \]
\[\lim_{y \rightarrow y_0} K(y,t) = 1\!\!1_d \frac{1}{(4 \pi t)^{
3/2}} \left[1-\frac{t}{2 a^ 2}+\sum_{n=1}^{\infty} (-1)^n \exp \left(-
\frac{\pi^ 2 a^2 n^ 2}{t}\right) \left(2-t-4 \pi^2 n^2 \frac{a^2}{t} \right)
\right], \ \ S^3, \]
where $d$ is the dimension of the Dirac algebra, in our case $d=4$.

\newpage
%\vskip1cm
\noindent{\large\bf Figure captions}
\bigskip

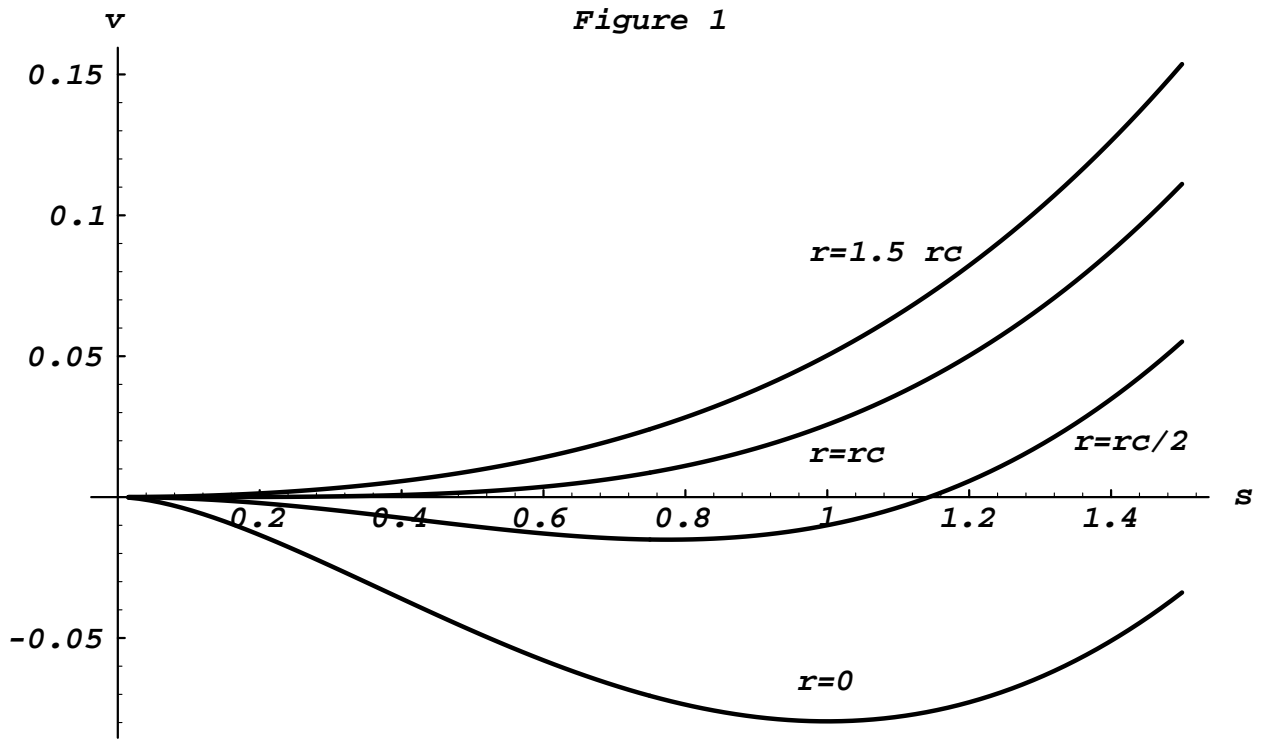
\noindent{\bf Fig. 1.} In this figure, $v \equiv V a^2$, $s \equiv \sigma
a$ and
$r \equiv \frac{R}{M^2}$. One sees that for $r>rc \equiv \frac{R_{cr}}
{M^2}$ the symmetry is restored in $S^2$.
\medskip

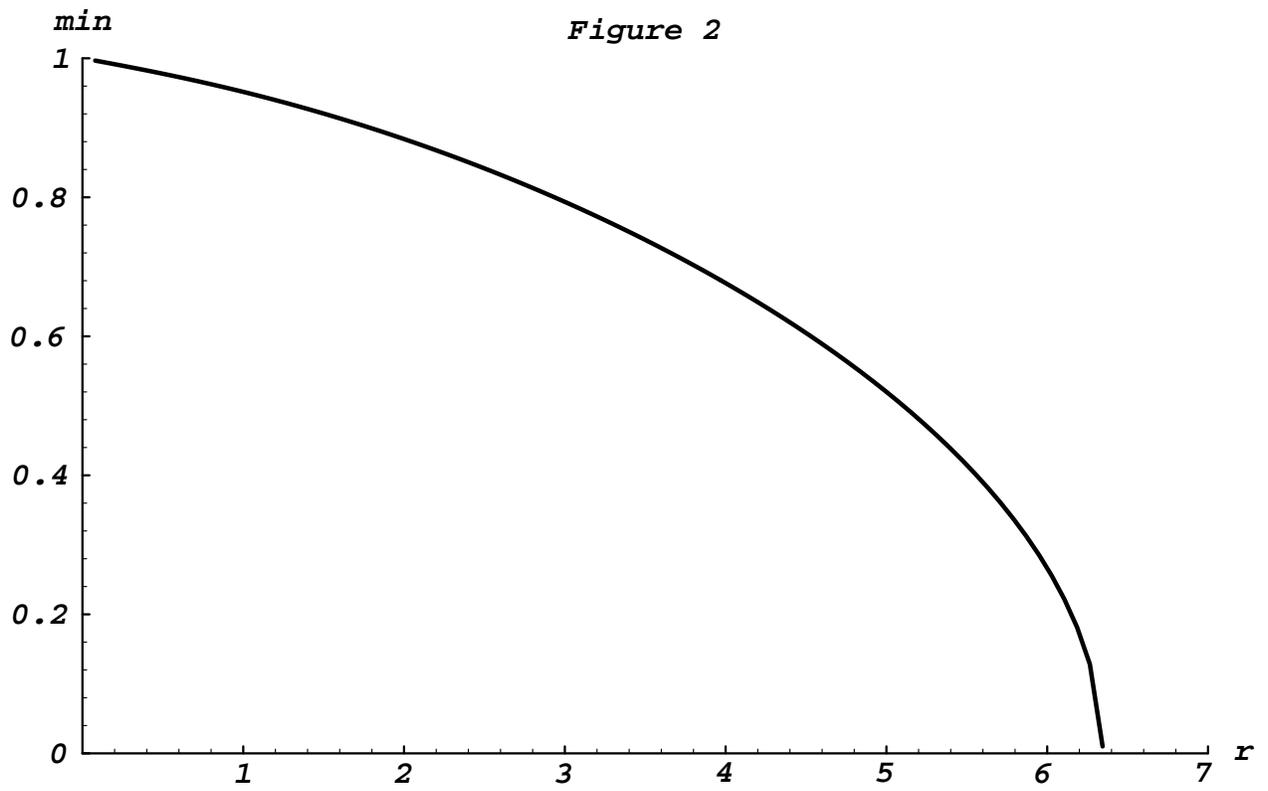
\noindent{\bf Fig. 2.} In this figure, $r \equiv \frac{R}{M^2}$ and ${\rm min}
=a \sigma_{min}$, where
$\sigma_{min}$ is the value of $\sigma$ at which $V$ attains its
minimum. The continuous character of the transition
which takes place in $S^2$ is clearly exhibited.
\medskip

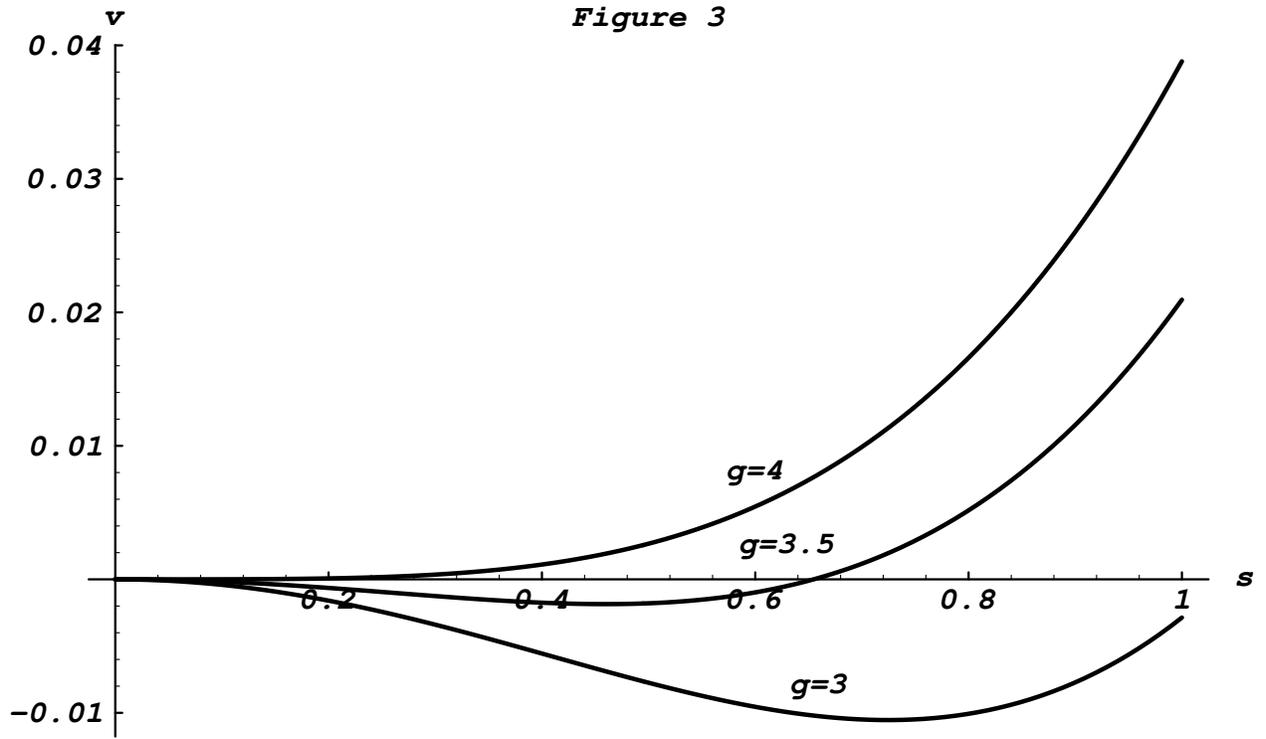
\noindent{\bf Fig. 3.} Here, $v \equiv V a^3$, $s \equiv \sigma a $ and $g
\equiv \frac{-\lambda}{a}$.
It is clearly seen that there may be a second order phase transition in
$S^3$ and
that it takes place when $g=4$.
\medskip

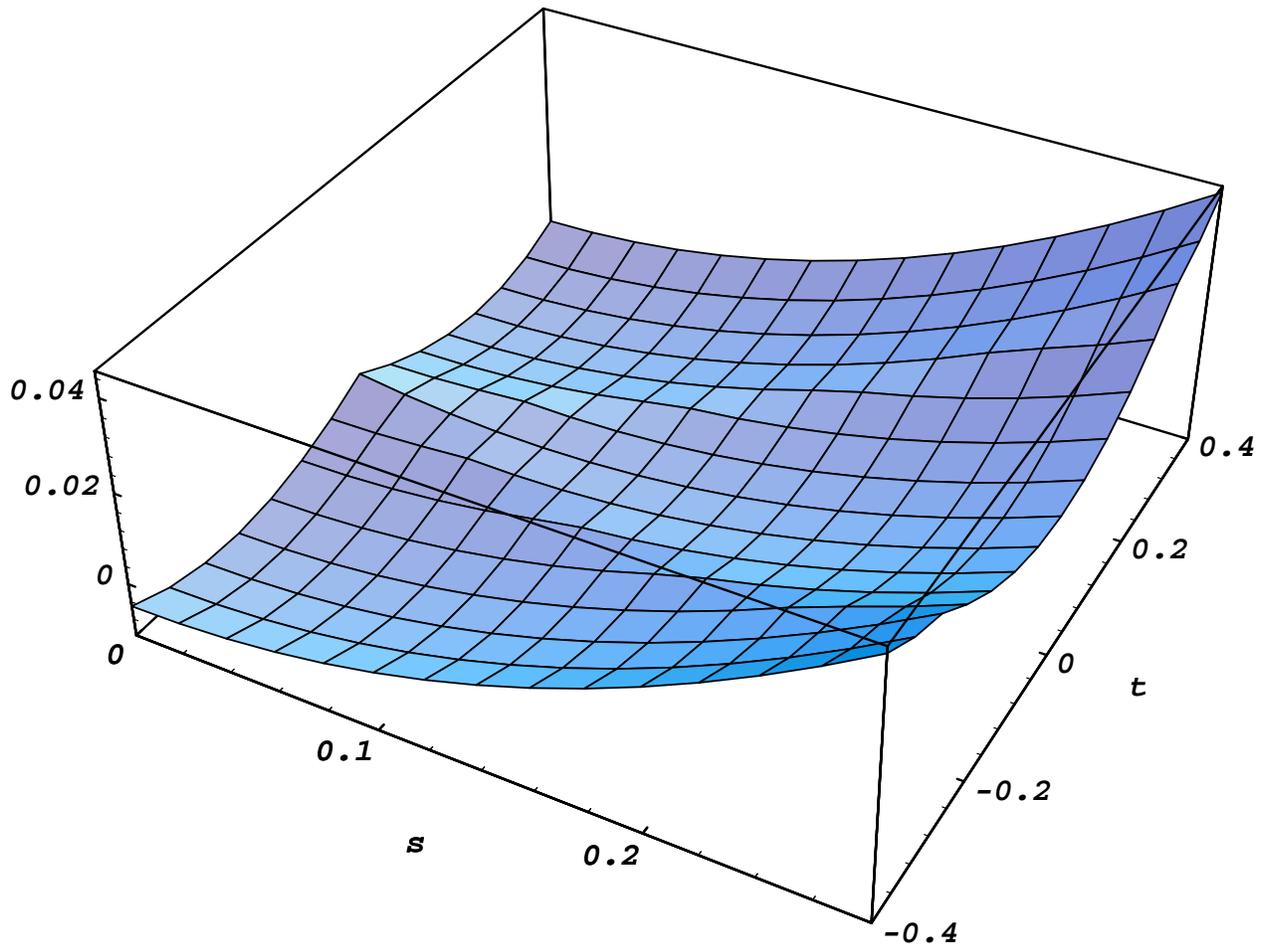
\noindent{\bf Fig. 4.} This figure is a plot of $V a^3$
in $H^3$ in terms of $s \equiv
\phi a$ and $t\equiv \chi a$. We have taken $\frac{a}{\lambda}=1$
and $\frac{a}{\kappa}=0.25$, so $\frac{1}{\lambda}-\frac{1}{
\kappa}>0$ and Z$_2$ is broken.
\medskip

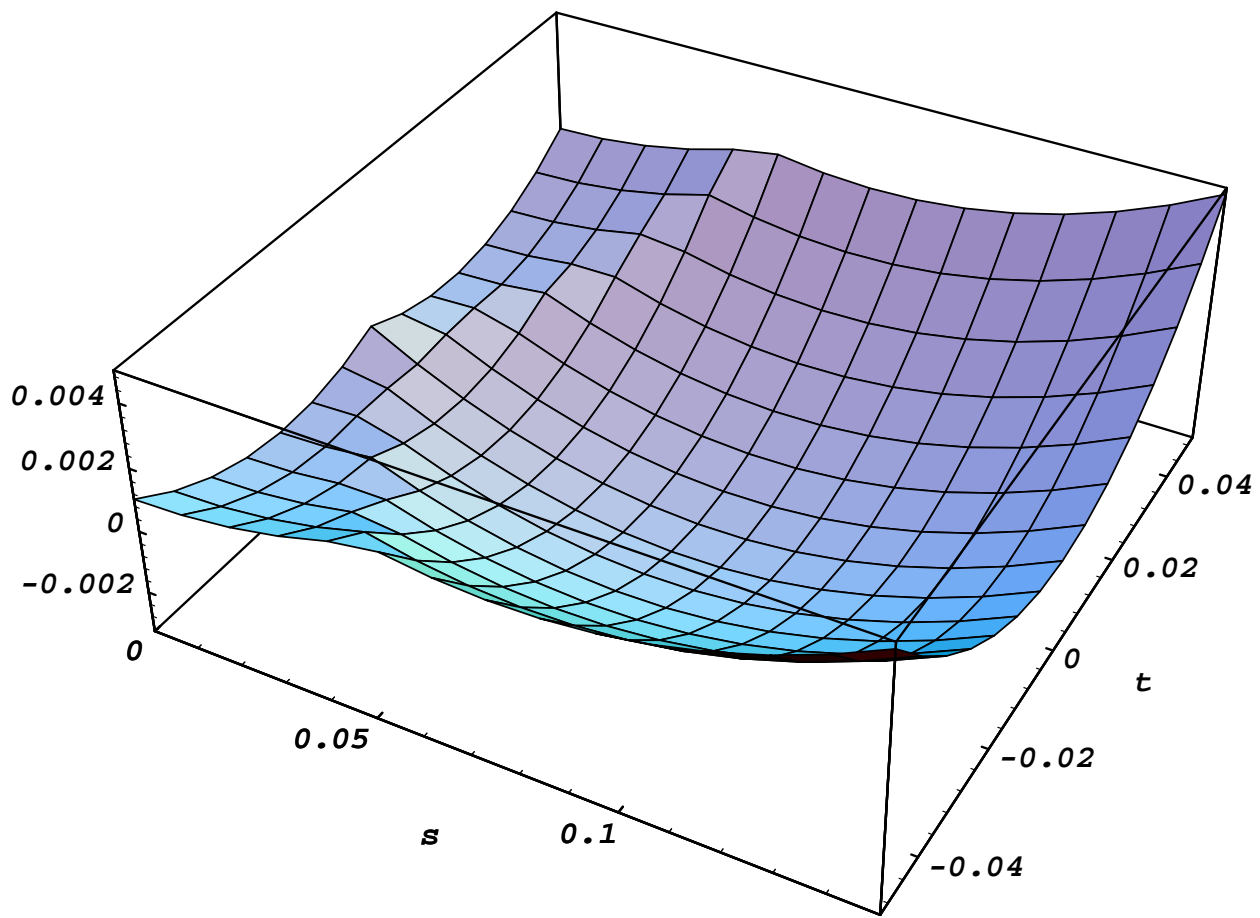
\noindent{\bf Fig. 5.} This is also a plot of $V a^3$ in $H^3$,
with $s$
and $t$ defined as in Fig. 4.
Here we have $\frac{1}{\lambda}-\frac{1}{\kappa}<0$
and P is broken.
\medskip

\end{document}